\documentstyle[prd,aps,manuscript,epsfig]{revtex}
\newcommand{\ua}{\sigma(^{\uparrow})}
\newcommand{\uu}{\sigma(^{\uparrow \uparrow})}
\newcommand{\da}{\sigma(^{\downarrow})    }
\newcommand{\ud}{\sigma(^{\uparrow \downarrow})  }
\newcommand{\dd}{\sigma(^{\downarrow \downarrow}) }
\newcommand{\du}{\sigma(^{\downarrow \uparrow})    }
\newcommand{\dss}{\Delta \sigma^{s} }
\newcommand{\dws}{\Delta \sigma^{d}}
\newcommand{\gev}{\,{\rm GeV}}
\newcommand{\tev}{\,{\rm TeV}}
\newcommand{\gvs}{\,{\rm (GeV}/c)^{2}}
\begin{document}
\begin{flushright}
Communication JINR E2-97-78
\end{flushright}
\vspace{2cm}
\title{Structure of  spin-dependent scattering amplitude \\
       and spin effects at small angles at  RHIC energies}
 \author{N. Akchurin}
\address{
Department of Physics and Astronomy, University of Iowa,
Iowa City, IA 52242, U.S.A. }
\author{S. V. Goloskokov, O. V. Selyugin }
\address{BLTPh, Joint Institute for Nuclear Research,
 141980 Dubna, Moscow Region, Russia}
\date{\today}
\maketitle
\begin{abstract}
Spin-dependent pomeron effects are analyzed for elastic $pp$
scattering and calculations for spin-dependent differential cross sections,
analyzing power and double-spin correlation parameters are carried out for
the energy range of the Relativistic Heavy Ion Collider (RHIC) at BNL.
In this energy range, $50 \leq \sqrt{s} \leq 500 \ \gev$, the structure of
pomeron-proton coupling can be measured at RHIC with colliding polarized
proton beams.
\end{abstract}

\pacs{13.85.Dz, 12.40.Nn, 13.88.+e}

\section{Introduction}

The spin structure of the pomeron is still an unresolved question
in the diffractive scattering of polarized particles.
There have been many observations of spin effects at high energies and
at fixed momentum transfers \cite{krish,nur}; several
attempts to extract the spin-flip amplitude from the experimental
data  show that the ratio of spin-flip to spin-nonflip
amplitudes can be non-negligible and may be independent of energy
\cite{akch,sel-pl}.  In all of these cases, the pomeron exchange is
expected to contribute the
observed spin effects at some level.
The large rapidity events that are observed at CERN \cite{ua8} and
DESY \cite{h1}, and all diffractive and elastic high energy reactions are
predominated by pomeron exchange, thus making the pomeron a popular field of
study.
Extensive polarized physics programs
are proposed at HERA, RHIC and LHC (see {\it e.g.}\cite{bunce,now,gur})
in order to
shed light on these and other spin effects in hadron reactions.

It is generally believed, based on calculations of the simplest QCD diagrams,
that the spin effects diminish as inverse power of center-of-mass energy, and
that the pomeron exchange does not lead to appreciable spin effects in
the diffraction region at super-high energies.
Complete calculations of the full set of helicity scattering amplitudes
in diffraction region cannot be carried out presently since they
require extensive treatment of confinement and contributions from many
diagrams.  Semi-phenomenological models, however, have been developed
with parameters which are expected to be fixed with the aid of data from
experiments \cite{soff,gol}.

Vacuum $t$-channel amplitude is usually associated with
two-gluon exchange in QCD \cite{low}. The properties of the spinless pomeron
 were analyzed on the basis of a QCD model,
by taking into account the non-perturbative properties of the theory
\cite{la-na,don-la}.  We refer to this as the
standard-pomeron model in this paper.

Some models predict non-zero spin effects as $s \to
\infty$ and $|t|/s \to 0$ limit.
   In these studies, the spin-flip amplitudes which lead to weakly altered
spin effects with increasing energy are connected with the structure of
hadrons and their interactions at large distances \cite{soff,gol}. In
reference \cite{soff}, the spin-dependence of the pomeron term is
constructed to model rotation of matter inside the proton. This approach is
based on Chou and Yang's concept of hadronic current density \cite{c-y}.
The model developed in reference \cite{gol} considers the
contribution of a sea quark-antiquark pair to hadron interactions at large
distances.

This picture can be related with the spin effects determined
by higher-order  $\alpha_{s}$ contributions in the framework of PQCD.
Really, it has been shown in the framework of QCD analysis at fixed
momentum transfer that these corrections to the simple two-gluon exchange
\cite{gol1} may lead to the spin-flip amplitude growing as $s$ in the
limit of $s \to \infty$. The similar effects can be determined by the
quark loops contributions \cite{gol2}

The high energy two-particle amplitude determined by pomeron
exchange can be written in the form:
\begin{equation}
T(s,t)=is I\hspace{-1.6mm}P(s,t) V_{h_1h_1I\hspace{-1.1mm}P}^{\mu} \otimes
V^{h_2h_2 I\hspace{-1.1mm}P}_{\mu}.    \label{tpom}
\end{equation}
Here  $I\hspace{-1.6mm}P$ is a function caused by the pomeron,
with a weak energy dependence $\sim (\ln{s})^n$; and
$V_{\mu}^{hhI\hspace{-1.1mm}P}$ are the pomeron-hadron vertices.
The perturbative calculation of the pomeron coupling structure
is rather difficult and the non-perturbative contributions are important
for momentum transfers of a few $\gvs$.

The calculation of Eq.(\ref{tpom}) in the non-perturbative
two-gluon exchange model \cite{la-na} and in the BFKL model \cite{bfkl}
shows that the pomeron couplings have the following simple form:
\begin{equation}
V^{\mu}_{hh I\hspace{-1.1mm}P} =\beta_{hh I\hspace{-1.1mm}P}\; \gamma^{\mu}.
\label{pmu}
\end{equation}
In this case, the pomeron contribution leads to a weak energy
dependence of the differential cross section with parallel
and antiparallel spins, and their difference drops as inverse power of $s$,
leading us to conclude that the spin effects are suppressed as a power of $s$.

This situation  changes dramatically when large-distance loop
contributions are considered which lead to a more complicated spin structure
of the pomeron coupling.  As mentioned above,
these effects can be determined by the hadron wave
function for the pomeron-hadron couplings, or by the gluon-loop
corrections for the quark-pomeron coupling \cite{gol-pl}.  As a result,
spin asymmetries appear that have weak energy dependence as $s \to
\infty$. Additional spin-flip contributions to the quark-pomeron vertex
may also have their origins in instantons, {\it e.g.} \cite{do,fo}.

In the framework of the perturbative QCD, the analyzing power of hadron-hadron
scattering was shown to be
of the order:
$$  A_N \ \propto \ m \alpha_s / \sqrt{p_{t}^{2}}$$
where $m$ is about a hadron mass \cite{ter}.  Hence, one would
expect a large analyzing power for moderate $p_{t}^{2}$
where the spin-flip amplitudes are expected to be
important for the diffractive processes.

In this paper, we examine the spin-dependent contribution of pomeron
to the differential cross sections with  parallel and antiparallel spins,
their possible magnitudes and energy dependence.  We also estimate
the possible experimental precisions for these observables in the RHIC energy
domain.

\section{The model amplitudes}

We use the standard helicity representation for
the hadron-hadron scattering amplitudes:
$ f_1=<++|M|++>$ and $ f_3=<+-|M|+->$, helicity nonflip amplitudes;
$ f_2=<++|M|-->$ and  $ f_4=<+-|M|-+>$, double-flip amplitudes; and
$ f_5 = <++|M|+->$, single-flip amplitude.
We assume, as usual, that at high energies
the double-flip amplitudes are small with respect
to the spin-nonflip one, $f_2(s,t) \sim f_4(s,t) \ll f_1(s,t)$
and that spin-nonflip amplitudes are approximately equal,
$f_{+}(s,t) =f_1(s,t) \sim f_3(s,t)$.  Consequently, the
observables are determined by two amplitudes: $f_{+}(s,t)$ and
$f_{-}(s,t)=f_5(s,t)$.  These customary
assumptions are also made for the models developed in \cite{soff,gol}.

We use the below normalization for the differential cross section:
$$ \sigma_0 = \frac{d\sigma}{dt} = \frac{4 \pi}{s(s-4m^2)}
       (|f_{+}|^2+2|f_{-}|^2), $$
and the analyzing power and the double spin correlation parameters are:
$$ \sigma_0 \ A_N = \frac{-8\pi}{s(s-4m^2)} Im[f_{-}^{*} f_{+}], $$
$$ \sigma_0 \ A_{NN} = \frac{4\pi}{s(s-4m^2)} 2 |f_{-}|^2.$$

The measured spin-dependent differential cross sections can be written in
the form:
\begin{eqnarray}
  N(^{\uparrow \uparrow})/{\cal L} = \uu & =
                         & (\sigma_0 /4) [1+A_N(P_1+P_2) \cos{\phi}
         +A_{NN}P_1 P_2 \cos^2 {\phi}], \\ \nonumber
  N(^{\downarrow \downarrow})/{\cal L} = \dd &=
                         & (\sigma_0 /4) [1-A_N(P_1+P_2) \cos{\phi}
         +A_{NN}P_1 P_2 \cos^2 {\phi}], \\ \nonumber
  N(^{\uparrow \downarrow})/{\cal L} = \ud &=
                         & (\sigma_0 /4) [1+A_N(P_1+P_2) \cos{\phi}
         -A_{NN}P_1 P_2 \cos^2 {\phi}], \\ \nonumber
  N(^{\downarrow \uparrow})/{\cal L} = \du &=
                         & (\sigma_0 /4) [1-A_N(P_1+P_2) \cos{\phi}
         -A_{NN}P_1 P_2 \cos^2 {\phi}], \\ \nonumber
  \sigma_0 &=& [N(^{\uparrow \uparrow}) + N(^{\downarrow \downarrow}) +
	      N(^{\uparrow \downarrow}) + N(^{\downarrow
\uparrow})]/{\cal L}.
\end{eqnarray}
${\cal L}$ is the luminosity and
$\sigma_{0}$ is the normalized differential cross section.
$P_1$ and $P_2$ refer to the degree of beam polarizations for the
first and second beams, respectively, and $\phi$ is the azimuthal
scattering angle.  The arrows indicate the transverse spin orientations of
 the interacting protons.

If we adapt the following notation:
$$ \ua = [N(^{\uparrow \uparrow}) + N(^{\uparrow \downarrow})]/{\cal L},\ \ \
 \da = [N(^{\downarrow \downarrow}) + N(^{\downarrow \uparrow})]/{\cal L},$$
$$ \uu = [N(^{\uparrow \uparrow})+ N(^{\downarrow \downarrow})]/{\cal L} ,\ \ \
 \ud = [N(^{\uparrow \downarrow}) + N(^{\downarrow \uparrow})]/{\cal L},$$
then, the analyzing power, $A_N$, and the double-spin correlation parameter,
$A_{NN}$, can be extracted from the experimental measurements;
\begin{eqnarray}
A_{N}=\frac{\sigma(^{\uparrow})-\sigma(^{\downarrow})}
{\sigma(^{\uparrow})+\sigma(^{\downarrow})}=
\frac{\dss   }{\sigma_0} =                       
 \frac{-2 Im (f_{-}^{*} f_{+})}{|f_{+}|^2 + 2 |f_{-}|^2},  \label{an}
\end{eqnarray}

\begin{eqnarray}
A_{NN}=\frac{\uu-\ud}{\uu+ \ud}=
\frac{\dws   }{\sigma_0} =
 \frac{2 |f_{-}|^2}{|f_{+}|^2 + 2 |f_{-}|^2}.  \label{ann}
\end{eqnarray}

Hereafter, $\dss$ and $\dws$ refer to the single- and double-spin
cross section differences.  We follow the model developed in \cite{gol} 
closely and extend
it further for the calculation of spin-dependent differential cross sections.
Pomeron-proton coupling $V^{\mu}_{ppI\hspace{-1.1mm}P}$
is primarily connected with the proton structure at
large distances and the pomeron-proton coupling looks like:
\begin{equation}
V_{ppI\hspace{-1.1mm}P}^{\mu}(p,t)=m p_{\mu} A(t)+ \gamma_{\mu} B(t),
\label{prver}
\end{equation}
where $m$ is the proton's mass, $p$ is the hadron momentum and
$t$ is the four-momentum-transfer square.
$\gamma_{\mu} B(t)$ is a standard pomeron coupling
that determines the spin-nonflip amplitude.  The term $m p_{\mu} A(t)$ is
due to meson-cloud effects. This coupling leads to
spin-flip at the pomeron vertex which does not vanish in the
$s \to \infty$ limit. Using Eq.(\ref{prver}), 
we can calculate the spin-nonflip
and spin-flip amplitudes from the pomeron-proton vertex:
\begin{eqnarray}
|f_{+}(s,t)| \propto s \;|B(t)|, \nonumber\\
|f_{-}(s,t)| \propto m \; \sqrt{|t|}\; s\; |A(t)|. \label{fpm}
\end{eqnarray}
Hence,  $V^{\mu}_{ppI\hspace{-1.1mm}P}$ determine $A_{N}$ and
$A_{NN}$ parameters.
Both of the above amplitudes have the same energy dependence and
the ratio of spin-flip
to spin-nonflip amplitudes in this picture gives:
\begin{equation}
\frac{\;\;\;m\;|f_{-}(s,t)|}{\sqrt{|t|}\;|f_{+}(s,t)|} \simeq
\frac{\;\;\;m^2\;|A(t)|}{|B(t)|} \simeq
  0.05 \,\,\,{\rm to}\,\,\,
0.07 \;\;\; {\rm for} \;\;\;  |t| \sim 0.5 \ \gvs  \label{fr}
\end{equation}
which is consistent with other estimates \cite{akch}.

The amplitudes $A$ and $B$ have a phase shift caused
by the soft pomeron rescattering. As a result, the analyzing power
determined by the pomeron exchange
\begin{equation}
A_{N} \ \frac{d\sigma}{dt} = 2m \sqrt{|t|} \ Im (AB^{*}) \label{epol}
\end{equation}
and appears to have a weak energy dependence.

The model also takes into account the contributions of the Regge terms
to both $f_{+}$ and $f_{-}$ amplitudes. So, the scattering amplitudes
are
\begin{eqnarray}
  f_{\pm}(s,t) =i s[f_{\pm}^{as} +
    \frac{(c_1 - i c_2)}{\sqrt{s}} f^{reg}_{+}
    + \frac{c_3 (1+i)}{\sqrt{s}} f^{reg}_{-}]  \ =  \
	 i s [f_{\pm}^{p}  + f_{\pm}^{r}],  \label{fpmr}
\end{eqnarray}
where $f^{as}_{\pm}, \ f^{reg}_{\pm}  $ are  functions which weakly
depend on energy and $c_{i}$ are parameters.
The asymptotic terms $f^{as}_{\pm}$ and $f^{reg}_{\pm}$
were calculated in the framework of the model \cite{gol}.
The Born amplitudes in the form of Eq.(\ref{fpm}) are modified by
pomeron rescattering.  The Regge contributions, Eq.(\ref{fpmr}), were
represented in the simplest exponential form. These, and
some of the asymptotic function parameters, were obtained from the fit
to the experimental data (for details see \cite{gol}).
The model quantitatively describes all the known experimental data of the
proton-proton and proton-antiproton scattering, from
$\sqrt{s} = 10\ \gev$ up to $\sqrt{s} = 1.8\ \tev$ \cite{gol,slop}.
We thus expect the predictions for differential cross sections 
at RHIC energies to be
reliable.  We neglect a possible odderon contribution in these calculations.

\section{Spin correlation effects}

The meson-cloud \cite{gol} and the rotating matter current
\cite{soff} models quantitatively describe the experimental data on
elastic $pp$ scattering at fixed momentum transfers and can predict physical
observables (cross sections and asymmetries) at high energies.
The predictions, however, differ in size and sign for asymmetries.

We calculate the spin-dependent cross sections
using the above described amplitudes in Eq.(\ref{fpmr}),
and we use the parameters of the RHIC beams \cite{gur} for the estimation
of statistical errors.  We assume at
$\sqrt{s} \sim 200 \ \gev$, the luminosity is
${\cal L} = 3 \times 10^{31}$ cm$^{-2}$ sec$^{-1}$ 
and the average beam polarization per beam is $70\%$.  
The typical geometrical acceptance of the detector is taken to be $20\%$
and the running time is about a month.  The momentum transfer binning, 
we take
$\Delta t=0.05 \,\gvs$ at $|t| \leq 1.3 \,\gvs$ and
 $\Delta t=0.1 \,\gvs$ at $|t| \geq 1.3 \,\gvs$.

The energy dependence of the real-part of the nonflip amplitudes around
diffraction minimum ($Im f_{+}(s,t) \sim 0$), is model dependent
and may lead to different polarization predictions in this momentum
transfer range.

    The local dispersion relations have been used in \cite{gol}
 to determine the real-part  of $T(s,t)$.
 The model amplitude obeys the $s-u$ crossing symmetry
 which permits us to describe quantitativaly
 all the specific effects
 in the elastic proton-proton and proton-antiproton
 diffraction scattering in the wide energy region
 ($9.8 \ GeV \ \leq \sqrt{s} \leq  \ 1800  \ GeV$).
 For example, model
 discribe the difference of the differential cross sections for proton-proton
 and proton-antiproton scattering in domain of the diffraction minimum,
 the known polarization phenomena
 at $\sqrt{s}=9.8 \ GeV$ and $\sqrt{s}= 52.8 \ GeV$ and
 the experimental data of proton-antiproton scattering at
 $\sqrt{s} = 540$ and $630 \ GeV$.
 All there quantities are sensitive to the real part
 of the scattering amplitude.
 So, we can believe
 that $Re[T(s,t)]$ determined correctly and practically
 model-independent.

The differential
cross section calculations are shown in Fig. 1 a,b. The estimated errors are
statistical and are less than $1\%$.
At $\sqrt{s}=50 \ \gev$, the model described here
quantitatively reproduces the ISR data \cite{ISR}.
The diffraction minimum defined by the zero of the imaginary
part of the spin-nonflip amplitude is filled by the contributions of
the real-part of spin-nonflip and spin-flip amplitudes.
At $\sqrt{s}=120 \ \gev$, the dip becomes a $\it shoulder$
and increases by an order of magnitude when the energy goes from
$\sqrt{s}=120 \ \gev$ to $500 \ \gev$.
The model gives the same asymptotical predictions for
the  proton-proton and proton-antiproton differential cross sections.
Hence the cross section prediction at $\sqrt{s} = 500 \ \gev$
approximately corresponds to the  
$Sp \bar p S$ data at $\sqrt{s}= 540 \ \gev$.

Fig.2 a,b show calculations for $\dws$ as determined in Eq.(\ref{ann}).
It can be seen that the shape of $\dws$ is similar to the spin-averaged
differential cross sections. It has a sharp dip
and the magnitude of $\dws$ at the dip grows by an order of magnitude
in the energy interval $120 \leq \sqrt{s} \leq 500 \ \gev$.
Although the dip positions for differential cross sections and that
of $\dws$ seem to coincide at $\sqrt{s}=50 \ \gev$ but at higher
energies they move apart from each other since the dip of $\dws$ moves
more slowly.
The position of the $\dws$ minimum strongly depends on the model
parameters.  However, in the ranges of $|t| \simeq 0.7$ to $1.0 \gvs$ and
$|t| \geq 1.8 \gvs$, the results weakly depend on these parameters.
In the regions far from the dips,  both cross sections change
slowly, especially at $|t| \geq 2 \ \gvs$.

The energy dependence of $f_{+}(s,t)$ and $f_{-}(s,t)$ amplitudes
are determined in Eq.(\ref{fpm}) for the $V_{ppI\hspace{-1.1mm}P}$ vertex.
Hence, we have the same energy dependence for
$\dws$ and $\ud$ for the spin-pomeron models,
and the ratio of these quantities will be almost energy
independent, except around the diffraction minimum.

The calculated ratio of the spin-parallel and antiparallel
cross sections,  $R(s,t) = \uu / \ud $,
shows only a logarithmic dependence on energy in its first maximum
and is almost energy independent at $|t|=0.7 \ \gvs$, see (Fig. 3 a,b).
It is clear from these results that the spin effects can be sufficiently
large at all RHIC energies.


Relative errors in $R(s,t)$ are shown in Table I at two selected
momentum transfers and at $t_m$ where the first maximum of $R(t)$ is observed.
It is worthwhile to note that the errors with the growth of energy decrease
due to increase in the differential cross sections.
Also note that in all standard pomeron models this ratio is predicted to be
$$ R(s,t) = \uu / \ud \rightarrow 1 \ \ {\rm as}  \ \ s \rightarrow \infty.$$
However, the experiment data show a large deviation from unity \cite{krish}.
It is also interesting that at low-energies ($p_{L}=11.7$ and $18 \ \gev/c$)
\cite{p11,p18},  $R(s,t)$ does not change much
with energy at $|t| \sim 2 \ \gvs$ and is similar to our predictions
of $R(s,t)$ at the first maximum.

Our calculation for $A_N$ is shown in Fig. 4 a,b.
The magnitude and the energy dependence of this parameter depend on
the energy behavior of  the zeros 
of the imaginary-part of spin-flip amplitude
and the real-part of spin-nonflip amplitude.

The maximum negative values of $A_N$  coincide closely with
the diffraction minimum (see Figs. 1 and 4).  We find that the contribution
of the spin-flip to the differential cross sections is much less
than the contribution of the spin-nonflip amplitude in the examined
region of momentum transfers from these figures.  $A_N$ is determined
in the domain of the diffraction dip by the ratio
\begin{eqnarray}
                 A_N \sim Im f_{- } / Re f_{+}. \label{ir}
\end{eqnarray}
The size of the analyzing power changes from $-45\%$ to $-50\%$
at $\sqrt{s}=50 \ \gev$ up to $-25\%$ at $\sqrt{s}= 500 \ \gev$.
These numbers give the magnitude of the ratio Eq.(\ref{ir}) that does
not strongly depend on the phase between the spin-flip and
spin-nonflip amplitudes.  This picture implies that the diffraction minimum
is filled mostly by the real-part of the spin-nonflip amplitude
and that the imaginary-part of the spin-flip amplitude 
increases in this domain as well.

We observe that the dips are
different in speed of displacements with energy from Figs. 1 and 2.
In Fig. 4,  one sees that at larger momentum transfers,
$|t| \sim 2$ to $3 \ \gvs$, the analyzing power depends on energy
very weakly.

The expected errors for the analyzing power are small
in nearly all momentum transfer ranges examined.
They are summarized in Table II.


As pointed out before, the model \cite{soff} predicts a similar absolute
value but opposite sign for $A_{N}$ near the diffraction minimum.
The future $PP2PP$ experiment at RHIC \cite{gur}
should be able to provide data and help resolve these issues
on the mechanism of spin-effect generation at the
pomeron-proton vertex.

\section{Conclusion}

In the framework of the standard-pomeron model, the spin-flip
amplitude is defined only by the secondary Regge poles and the ratio
$$[\sigma(^{\uparrow \uparrow})-\sigma(^{\uparrow \downarrow})]/
\sigma(^{\uparrow \downarrow}) \propto 1/s$$
that rapidly decreases with growing $s$ due to the standard energy
dependence of the spin-flip amplitude.
If we drop the asymptotic term in the spin-flip amplitude from
Eq.(\ref{fpmr}) and keep only the second Regge terms 
that fall as $1/\sqrt{s}$,
we obtain the pre-asymptotic Regge contributions to the analyzing power
and the difference of the polarized cross sections (the results are shown
in Fig. 5 and 6).

The spin-flip amplitude of the second Regge contributions has
a large relative phase compared to the spin-nonflip amplitude of the pomeron.
Under this condition, too, the analyzing power can be large.
As one observes from Fig. 5, the spin-flip amplitude defined by the
Regge contributions can describe the experimental data at
$\sqrt{s}=23.4 \ \gev$ and give large effects at $\sqrt{s}=50 \ \gev$.
At higher energies, however, the effect quickly falls and becomes
insignificant.  Note that at $\sqrt{s} = 50 \ \gev$,  $A_N$ can be positive
and that its magnitude greatly depends on the size of the asymptotic term
which gives a negative contribution to the analyzing power.


If the spin-flip amplitude is determined by the standard-pomeron model,
its contribution will be clearly seen in experiments (see Figs 5, 6) 
if performed. In this case, the minimum in $\dws$ is
at small momentum transfers  $|t| \leq 0.7 \ \gvs$,
and is quickly shifted with growing energy.
In the ordinary dip region, there is a maximum in the difference
of the cross sections which falls as the inverse power of energy.
Moreover, at $|t|  \geq 2 \ \gvs$, in the spin-pomeron model,
$\sigma_{0}$ and $\dws$  do not appreciably change with energy
at fixed $t$ (Figs. 1,2).
The difference spin-dependent cross sections fall with energy
as quickly as the increasing momentum transfers (Fig. 6).

The energy dependence of cross sections
$\sigma(^{\uparrow \uparrow})$ and $\sigma(^{\uparrow \downarrow})$ can be
studied experimentally at RHIC. Note that significant spin effects
can have small relative errors at momentum transfers
$|t| \sim 2$ to $3 \gvs$
and direct information about
the nature of the spin-flip effects in the pomeron-proton coupling
can be obtained.
The future  $PP2PP$ experiment at RHIC should be able
to measure the spin-dependent cross section with parallel
$\sigma(^{\uparrow \uparrow})$ and antiparallel
$\sigma(^{\uparrow \downarrow})$ beam polarizations in
proton-proton scattering and the energy dependence of
the spin-flip and spin-nonflip amplitudes can be studied in the energy
range of RHIC, $50 \leq \sqrt{s} \leq 500\,\gev$.

The spin-structure of the pomeron couplings are determined by
the large-distance gluon-loop correction or by the effects of hadron wave
function.  Tests of the spin structure of QCD at large
distances can be carried out in diffractive reactions
in future polarized experiments at HERA, RHIC and LHC.

\acknowledgements

The authors are grateful to A. V. Efremov, W.-D. Nowak, S. B. Nurushev and
A. Penzo for fruitful discussions.


\newpage
\begin{table}
\caption{The expected relative errors for
$R(s,t)=\uu / \ud $ are tabulated below
at $|t_{1}|= 0.5 \ \gvs$, $|t_{2}|= 2.5 \ \gvs$ and
at $t_{m}$, where $R(s,t)$ is maximum.}
\vspace{2mm}
\begin{tabular}{r c c c}
$\sqrt{s} \ [\gev]$  & $\delta R(t_1)$ & $\delta R(t_m)$  & $\delta R(t_2) $   \\   \hline
 50        & $ 0.05\%$     & $ 5.5\%$       & $ 4.1\%  $   \\
 120       & $ 0.07\%$     & $ 2.2\%$       & $ 4.1\%  $   \\
 250       & $ 0.07\%$     & $ 1.1\%$       & $ 3.7\%  $   \\
 500       & $ 0.09\%$     & $ 0.6\%$       & $ 3.6\%$    \\
\end{tabular}
\end{table}

\begin{table}
\caption{The expected relative errors for $A_{N}$ are calculated
at $|t_{1}|= 0.5 \ \gvs$, $|t_{2}|= 2.5 \ \gvs$ and
at $t_{m}$, where $A_N$ is maximum.}
\vspace{.2cm}
\begin{tabular}{c c c c}
$\sqrt{s} \ [\gev]$ & $\delta A_{N}(t_1)$ & $\delta A_{N}(t_m)$  & $\delta A_{N}(t_2)
$   \\   \hline
 50        & $ 3.1\%$     & $ 3.7\%$       & $ 16\%  $   \\
 120       & $ 2.3\%$     & $ 2.0\%$       & $ 20\%  $   \\
 250       & $ 3.5\%$     & $ 1.4\%$       & $ 18\%  $   \\
 500       & $ 12.4\%$     & $ 1.1\%$       & $ 17\%$    \\
\end{tabular}
\end{table}

\newpage
\begin{figure}
\caption{
a) The calculated differential cross sections of the $pp$ elastic
scattering at $s^{1/2} = 50 \,\,{\rm GeV}$ (solid curve) and
at $s^{1/2} = 120 \,\,{\rm GeV}$ (dashed curve);
b)
$s^{1/2} = 250 \,\,{\rm GeV}$ (dotted curve) and
at $s^{1/2} = 500 \,\,{\rm GeV}$ (dashed curve).  The energy range between
50 GeV and 500 GeV correspond to the entire RHIC energy for polarized
protons.
}
\end{figure}

\begin{figure}
\caption{
a, b)
The calculated difference of the double spin differential cross sections at
the RHIC energy range.
The error bars indicate possible statistical errors for a realistic experiment
at RHIC.}
\end{figure}

\begin{figure}
\caption{
a, b)
The calculated ratio of the cross sections with parallel
and antiparallel spins at the RHIC energy range.
The error bars indicate possible statistical errors 
for a realistic experiment at RHIC.
}
\end{figure}

\begin{figure}
\caption{
The calculated analyzing power of the $pp$ elastic cross sections
for the weak energy dependence obtained in the model for the ratio
 $|F^{+-}(s,0)|/|F^{++}(s,0)| \propto f(\ln{s})$
a) at $s^{1/2} = 50 \,\,{\rm GeV}$ (solid curve) and
at $s^{1/2} = 120 \,\,{\rm GeV}$ (dashed curve);
b)
$s^{1/2} = 250 \,\,{\rm GeV}$ (dotted curve) and
at $s^{1/2} = 500 \,\,{\rm GeV}$ (dashed curve).
The error bars indicate expected statistical errors for a realistic experiment
at RHIC.
}
\end{figure}

\begin{figure}
\caption{
The analyzing power for the $pp$ elastic cross sections
for the rapid energy dependence of the ratio
 $|F^{+-}(s,0)|/|F^{++}(s,0)| \propto  s^{-1/2}$
 (the experimental data
at $P_L = 300 \,\,{\rm GeV}/c$ are represented by open circles) ;
the results of calculations are shown for
$P_L = 300 \,\,{\rm GeV}$ (solid curve);
at $s^{1/2} = 50 \,\,{\rm GeV}$ (long-dashed curve);
at $s^{1/2} = 120 \,\,{\rm GeV}$ (short-dashed curve);
and at $s^{1/2} = 500  \,\,{\rm GeV}$ (dotted curve).
}
\end{figure}

\begin{figure}
\caption{
The calculated difference of the double spin differential
for the rapid energy dependence of the ratio
 $|F^{+-}(s,0)|/|F^{++}(s,0)| \propto  s^{-1/2}$
  The center-of-mass energy span corresponds to the
RHIC energy and each curve shows the results at four different {\it cms}
energies as indicated on the figure.}
\end{figure}

\newpage
  \vspace*{.5cm}
\epsfxsize= 9.cm
\centerline{\epsfbox{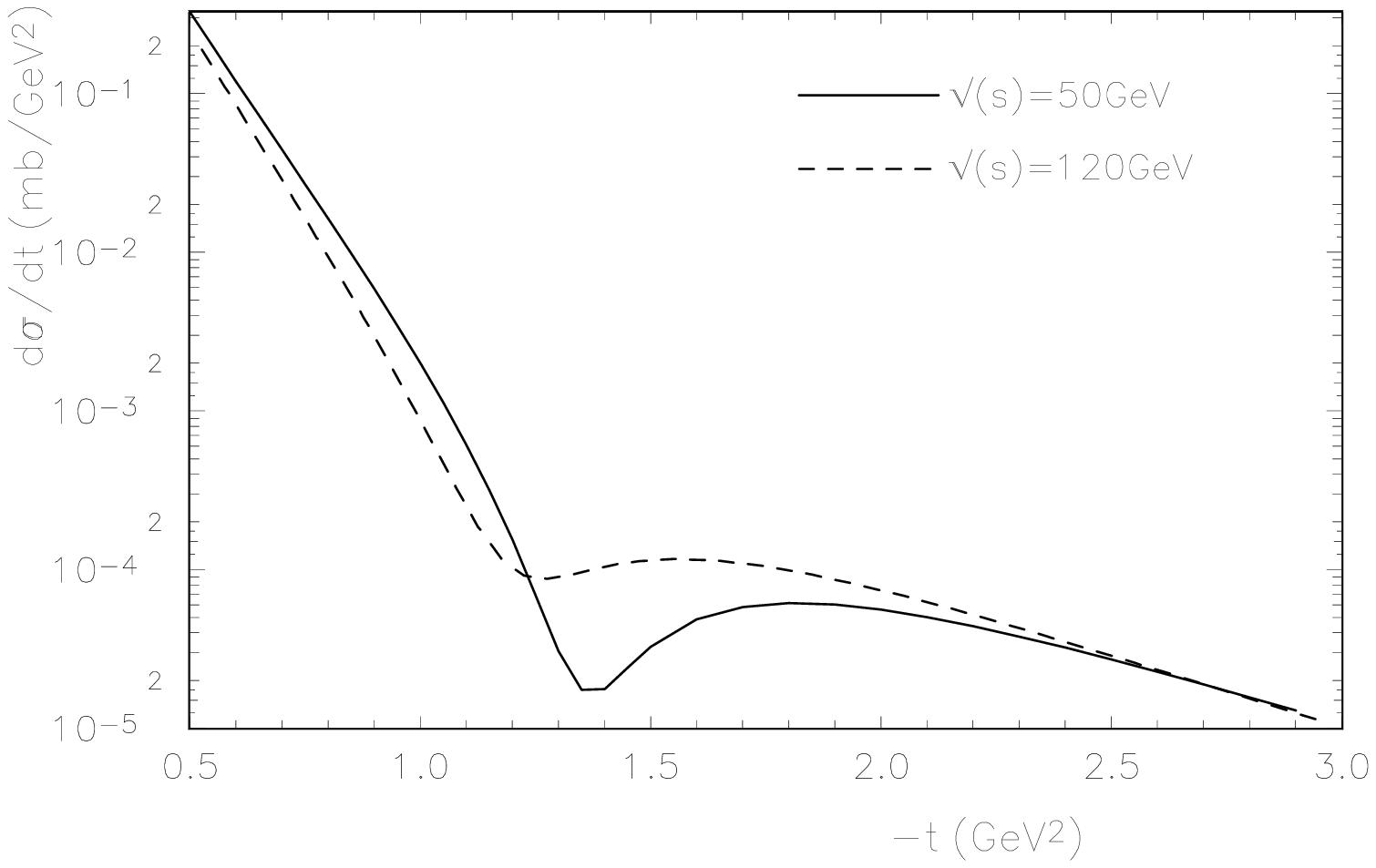}}
\begin{center}
Fig.1 a
\end{center}
%
\epsfxsize=9.cm
\centerline{\epsfbox{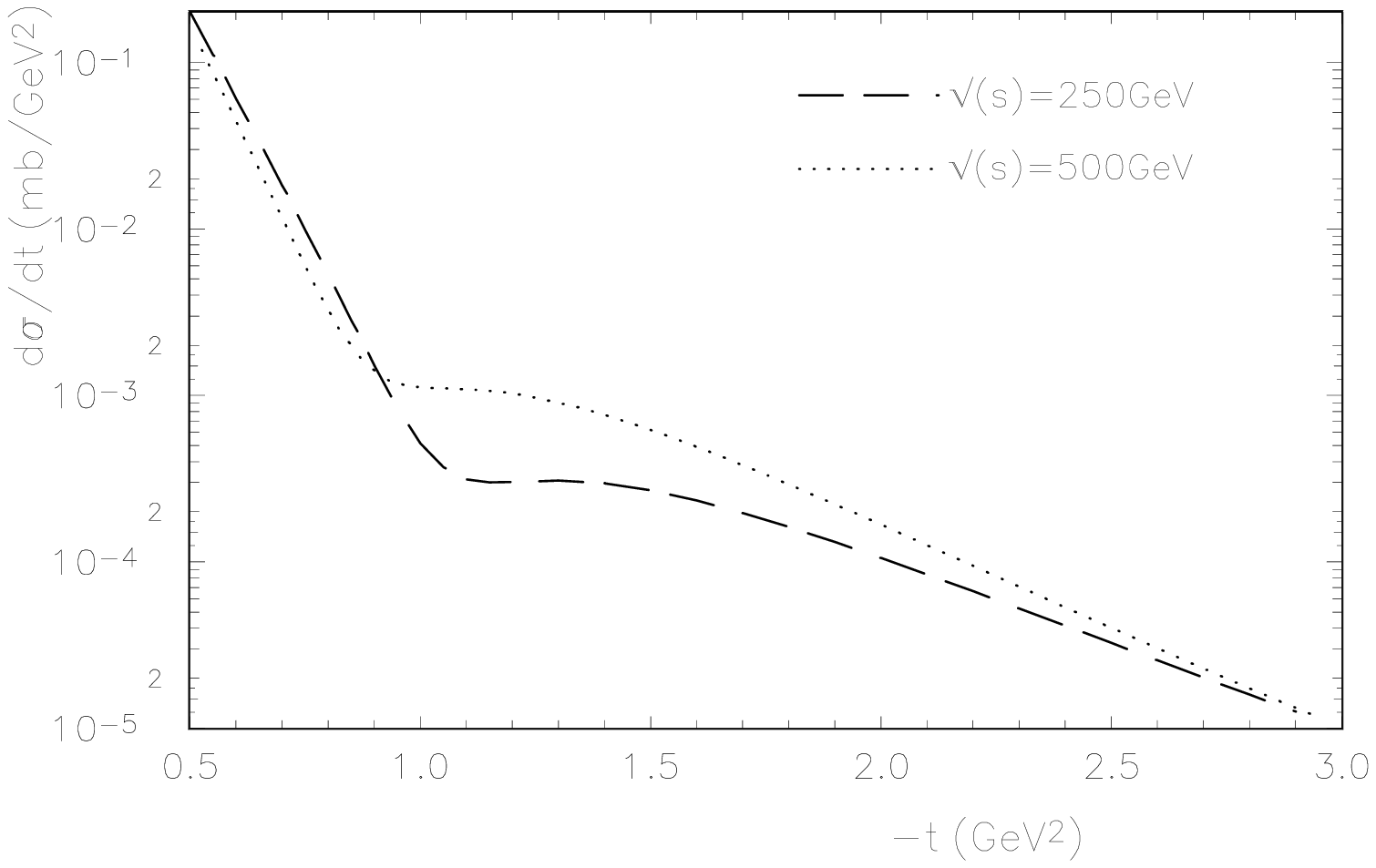}}
\vspace*{2cm}
\begin{center}
Fig.1 b
\end{center}
\newpage
%
  \vspace*{.5cm}
\epsfxsize= 9.cm
\centerline{\epsfbox{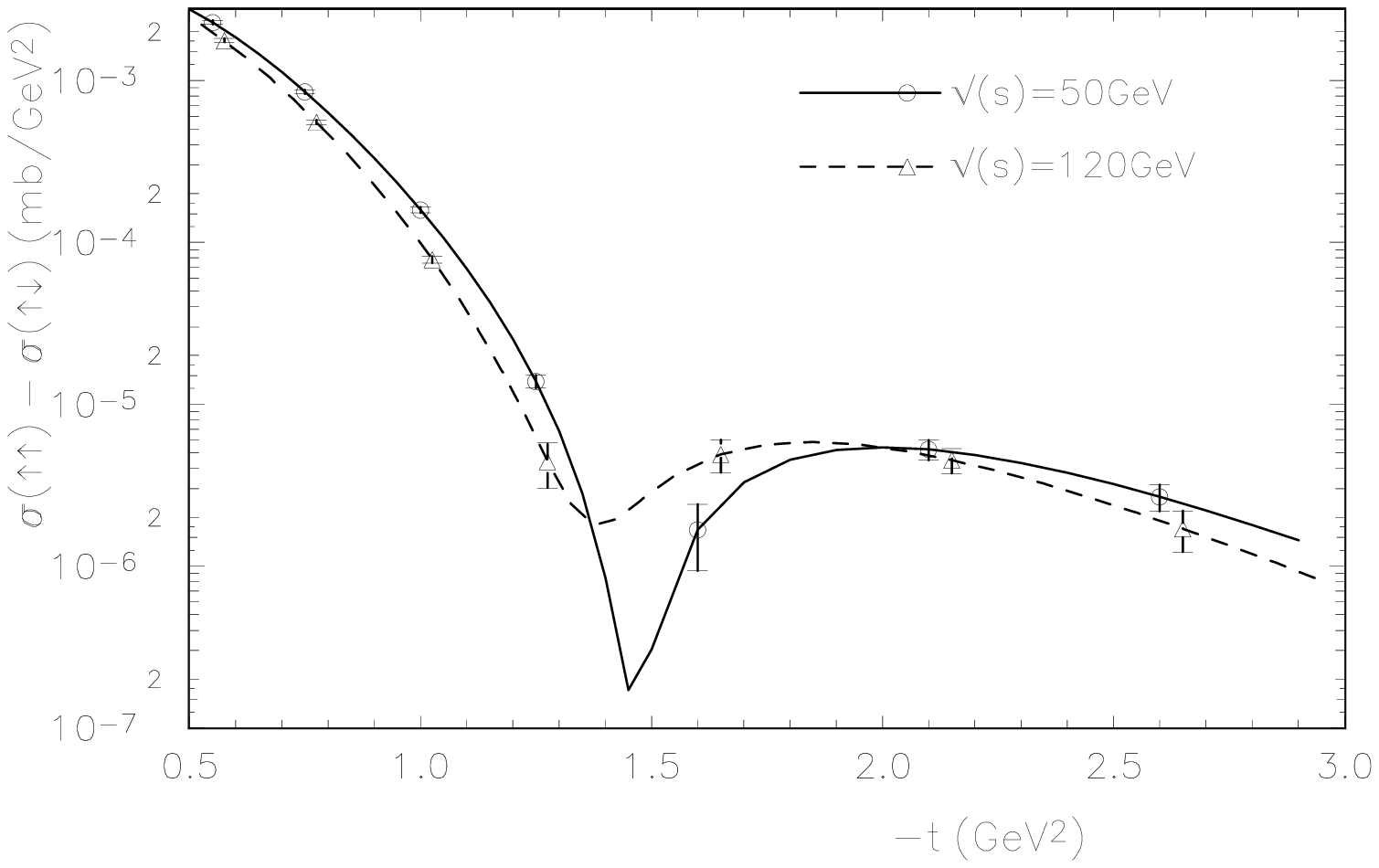}}
\begin{center}
Fig.2 a
\end{center}
%
\epsfxsize= 9.cm
\centerline{\epsfbox{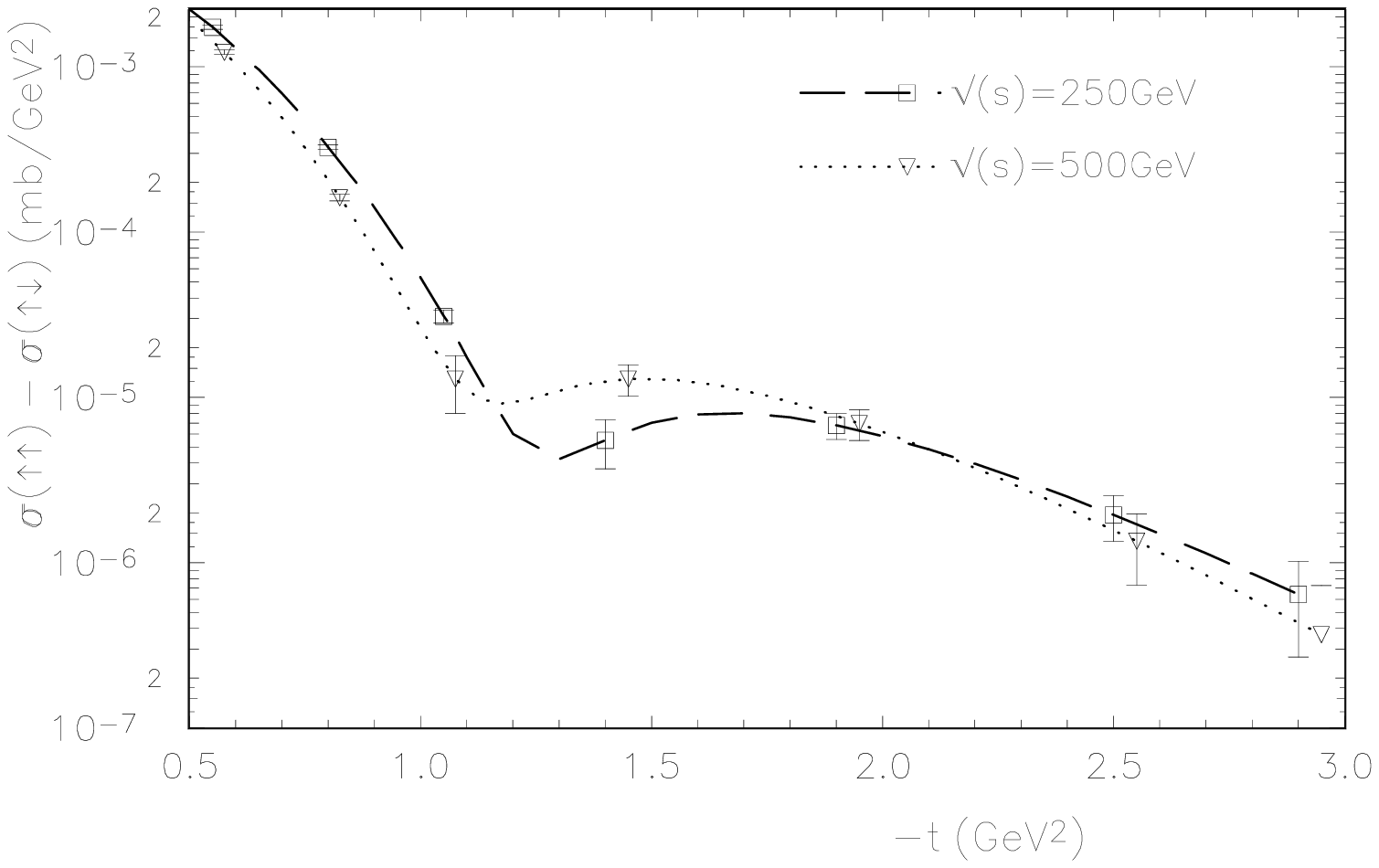}}
\vspace*{2cm}
\begin{center}
Fig.2 b
\end{center}
\newpage
%
  \vspace*{.5cm}
\epsfxsize=9.cm
\centerline{\epsfbox{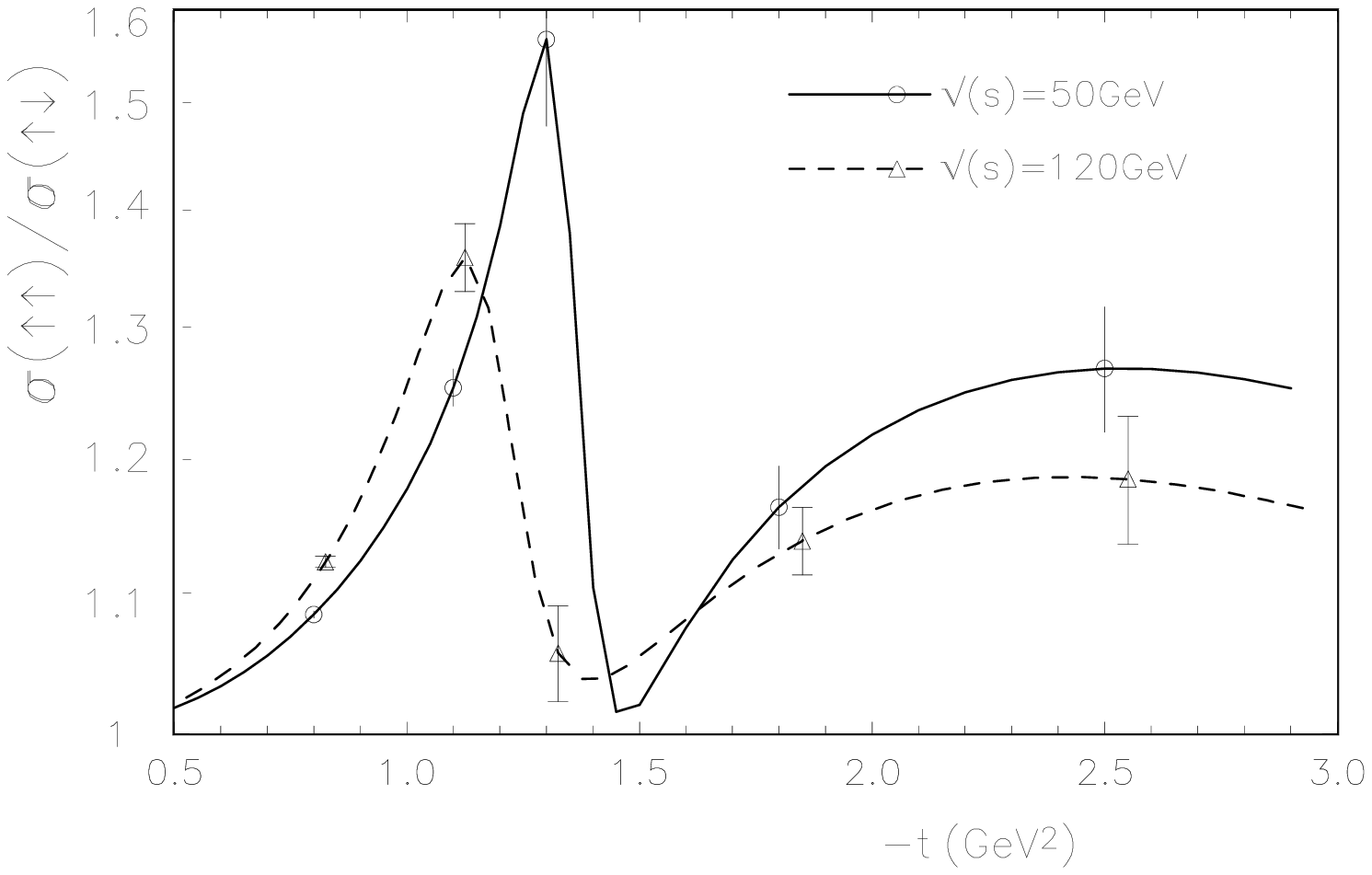}}
\begin{center}
Fig.3 a
\end{center}
%
\epsfxsize=9.cm
\centerline{\epsfbox{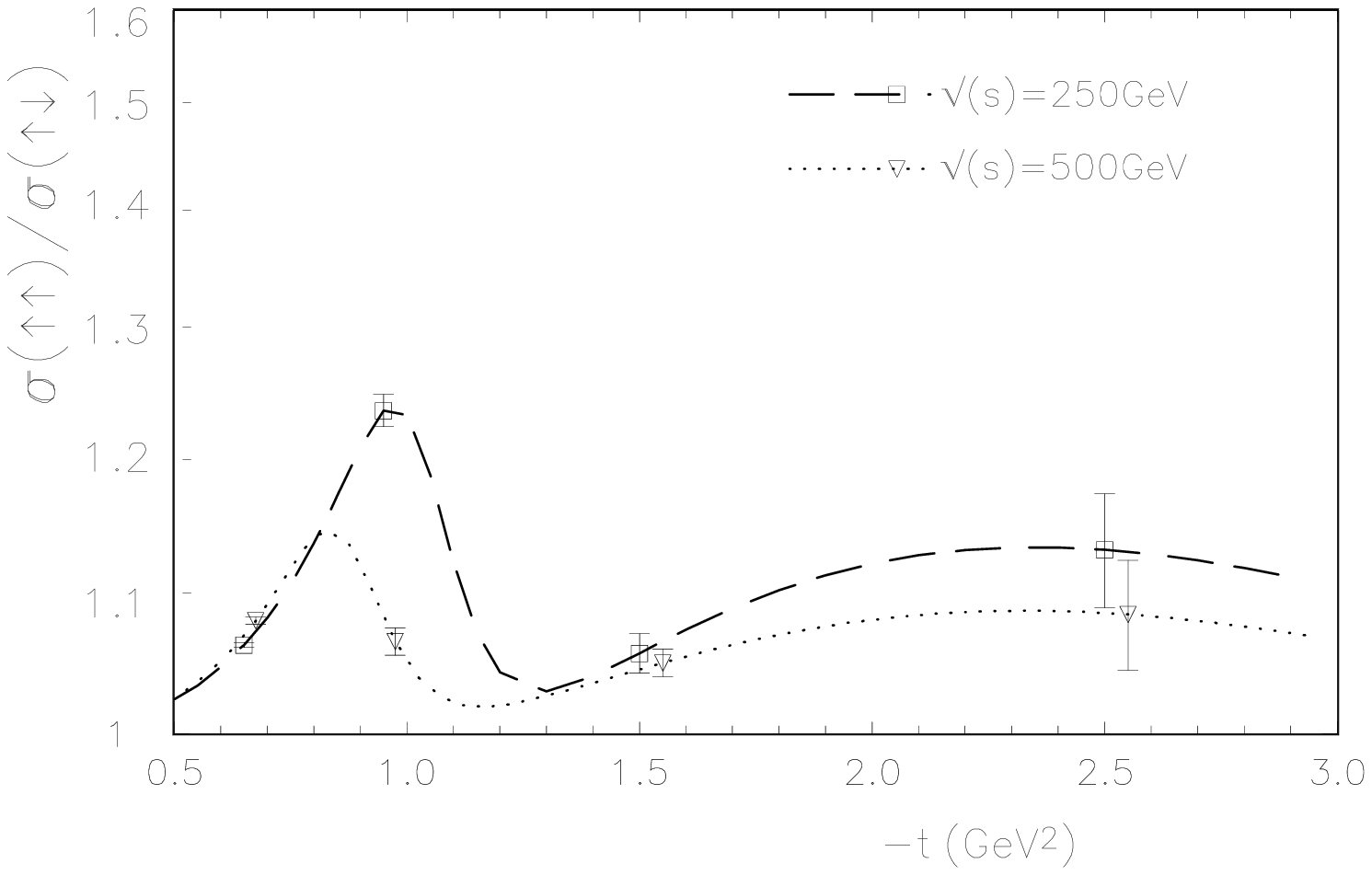}}
\vspace*{2cm}
\begin{center}
Fig.3 b
\end{center}
\newpage
%
  \vspace*{.5cm}
\epsfxsize=9.cm
\centerline{\epsfbox{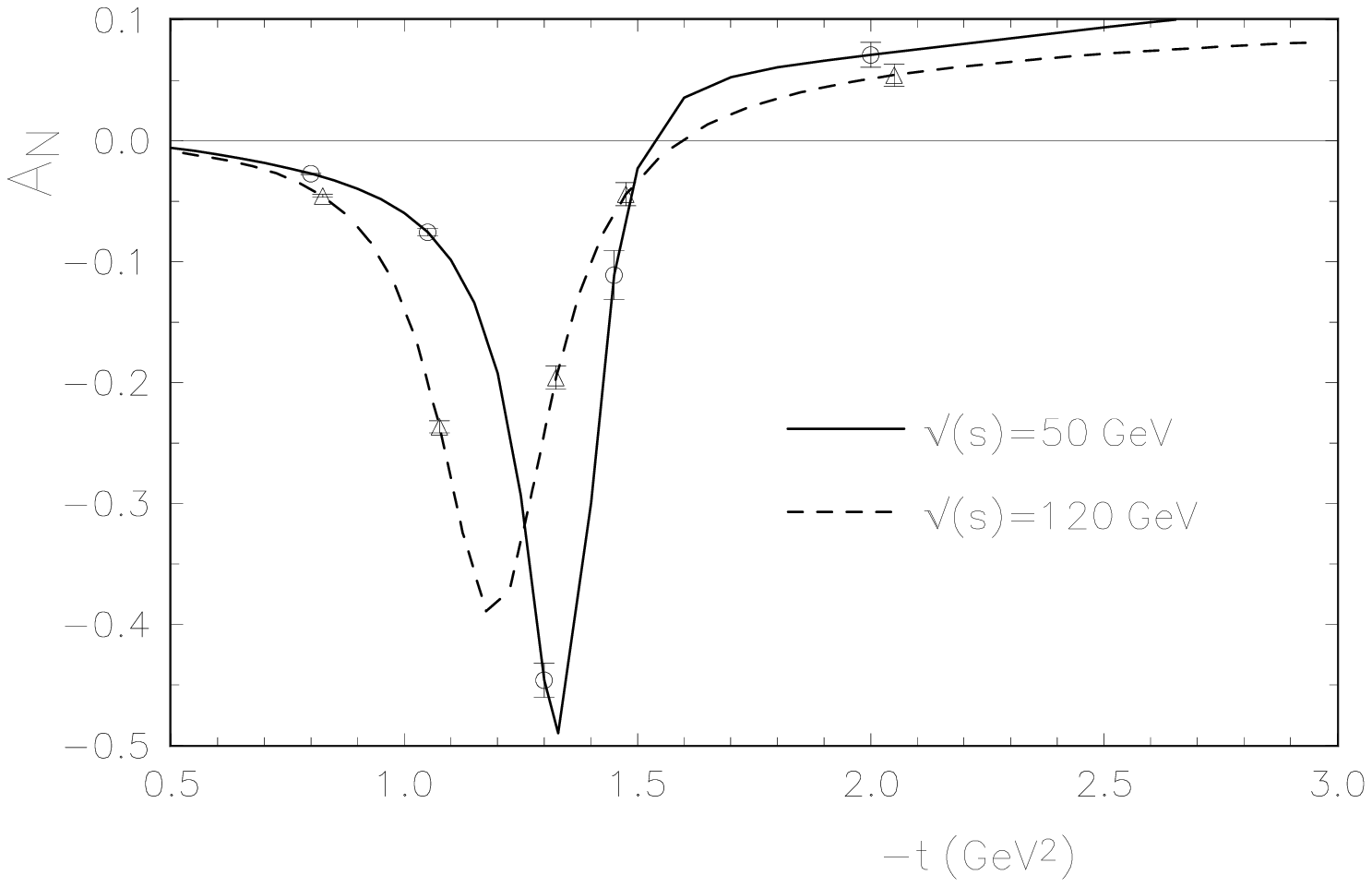}}
\begin{center}
Fig.4 a
\end{center}
%
\epsfxsize=9.cm
\centerline{\epsfbox{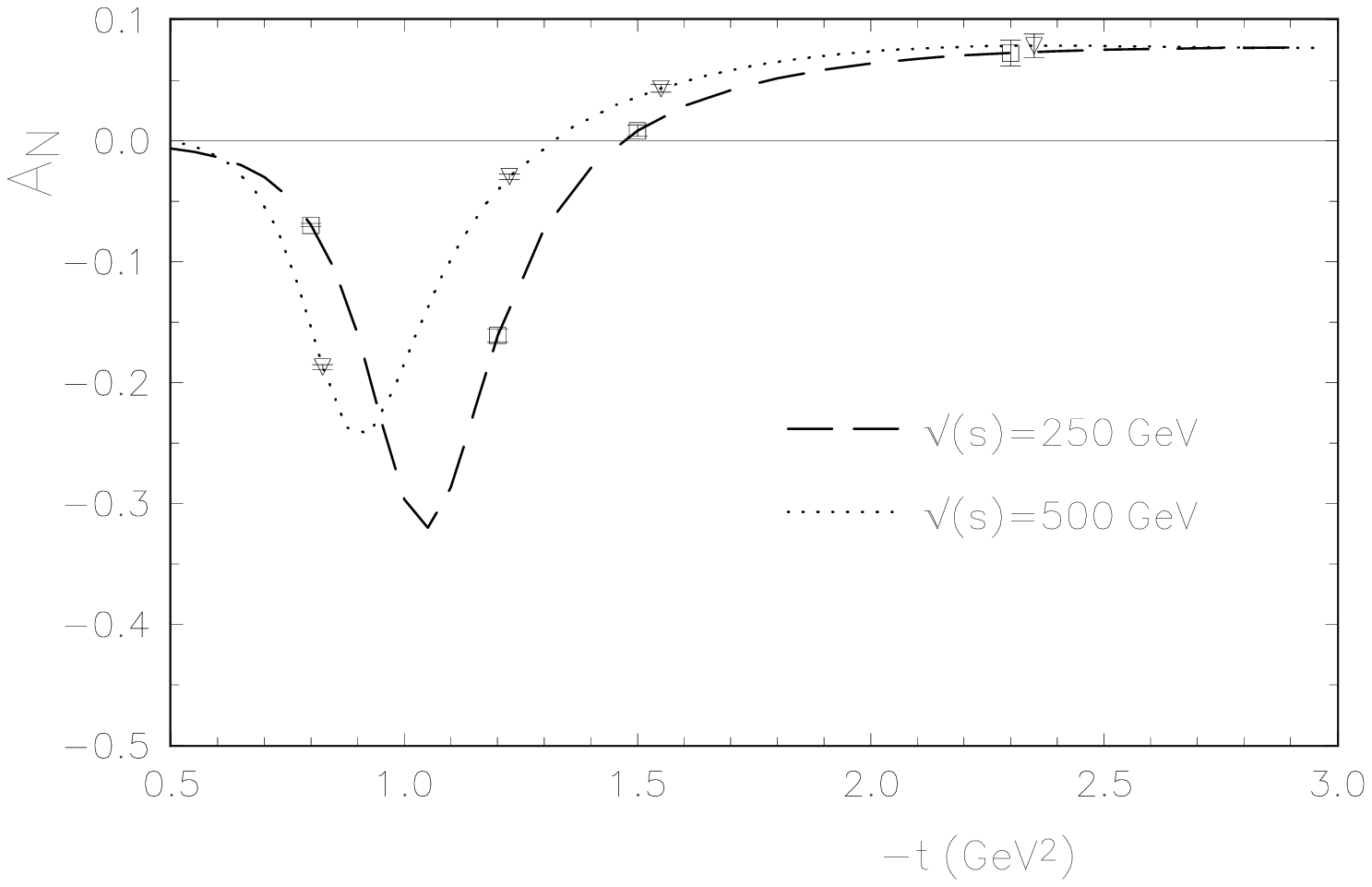}}
\vspace*{2cm}
\begin{center}
Fig.4 b
\end{center}
\newpage
%
  \vspace*{.5cm}
\epsfxsize=9.cm
\centerline{\epsfbox{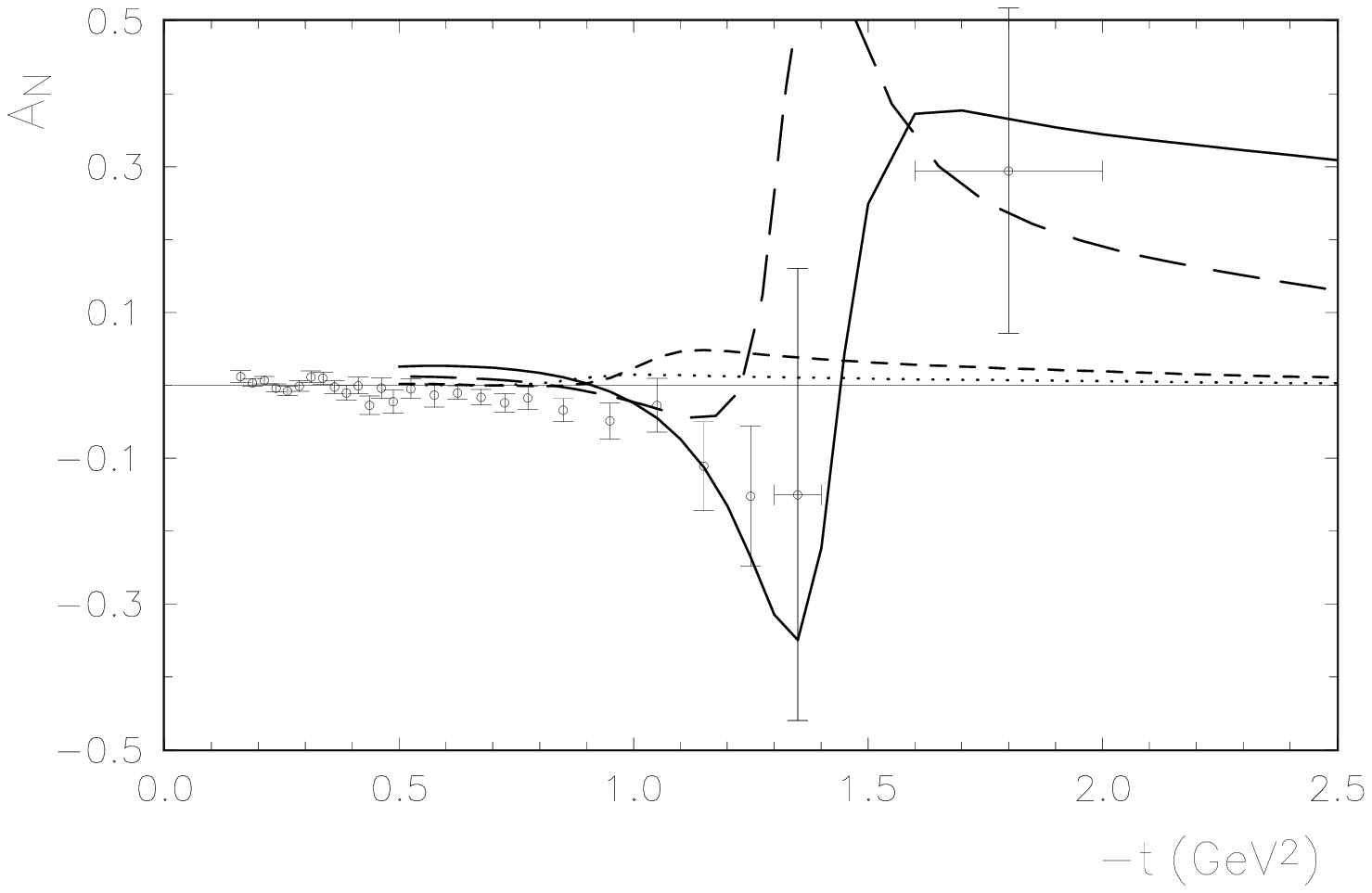}}
\begin{center}
Fig.5
\end{center}
%
\epsfxsize=9.cm
\centerline{\epsfbox{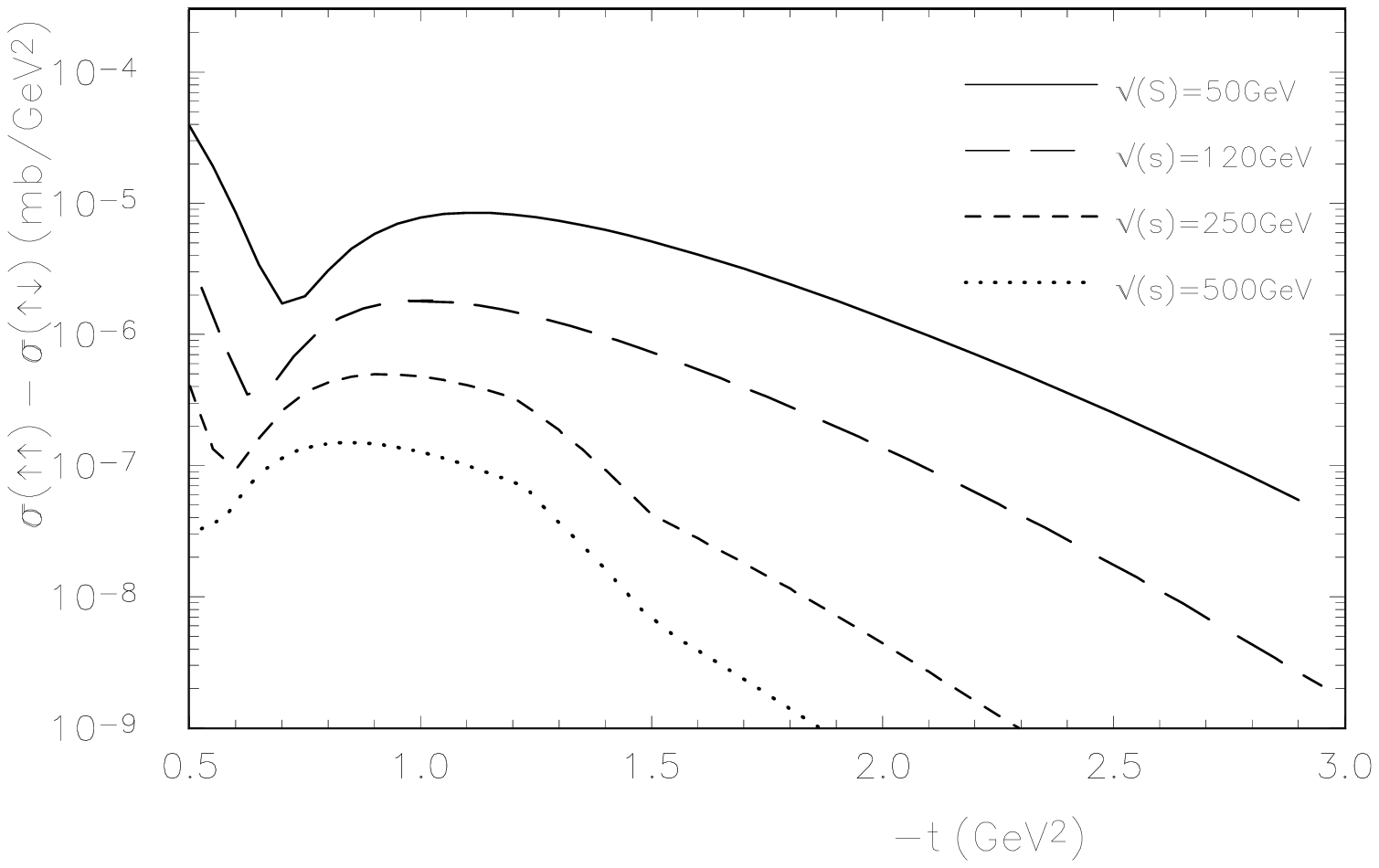}}
\vspace*{2cm}
\begin{center}
Fig.6
\end{center}

\end{document}